# Optimising Element Choice for Nanoparticle Radiosensitisers

*Stephen J McMahon[1,2]\*, Harald Paganetti[1], Kevin M Prise[2]*

1. Department of Radiation Oncology, Massachusetts General Hospital & Harvard Medical School

Boston, MA, 02114, USA

2. Centre for Cancer Research and Cell Biology, Queen's University Belfast, Belfast, BT9 7AE, UK

## *Abstract*

There is considerable interest in the use of heavy atom nanoparticles as theranostic contrast agents due to their high radiation cross-section compared to soft tissue. However, published studies have primarily focused on applications of gold nanoparticles. This study applies Monte Carlo radiation transport modelling using Geant4 to evaluate the macro- and micro-scale radiation dose enhancement following X-ray irradiation with both imaging and therapeutic energies on nanoparticles consisting of stable elements heavier than silicon. An approach based on the Local Effect Model was also used to assess potential biological impacts. While macroscopic dose enhancement is well predicted by simple absorption cross-sections, nanoscale dose deposition has a much more complex dependency on atomic number, with local maxima around germanium (Z=32) and gadolinium (Z=64), driven by variations in secondary Auger electron spectra, which translate into significant variations in biological effectiveness. These differences may provide a valuable tool for predicting and elucidating fundamental mechanisms of these agents as they move towards clinical application.

## Introduction

Radiotherapy's primary objective is to selectively deliver high doses of radiation to tumours while sparing surrounding normal tissues. Clinical progress has been driven in recent years by significant technical advances in radiation delivery, with advanced techniques such as Intensity Modulated Radiotherapy (IMRT) and Volumetric Arc Therapy (VMAT) enabling highly conformal radiation delivery. These techniques are increasingly coupled with image guidance, currently using single and dual energy kV X-rays and with interest into expanding these techniques to incorporate Linac-MRI approaches in the future.

Despite these advances, tumour dose escalation is often limited by the presence of nearby organs at risk, as inherent uncertainties in treatment delivery place strict limits on prescription doses to minimise radiotherapy-related side effects. As a result, there is a significant interest in techniques which further improve dose specificity to tumour volumes.

One approach to selectively spare healthy tissue is through the introduction of contrast agents – materials of high atomic number (Z) which strongly absorb ionising radiation. If these particles can be delivered preferentially to tumour volumes, they can selectively increase the target's absorption, offering both improved image contrast and dose conformality.

This approach has long been hampered by the lack of a suitable tumour-specific mechanism for delivering these contrast agents, but in recent years there has been significant interest in the application of high-Z nanoparticles for this purpose, following early work demonstrating the efficacy of gold nanoparticles as radiosensitising agents in mice (1). These nanoparticles were able to exploit the leaky tumour vasculature to achieve selective uptake in tumour volumes via the enhanced permeability and retention effect (EPR), and when combined with radiotherapy gave significant improvements in tumour control and overall survival in mice compared to radiotherapy alone.

Following this early work, there have been several hundred publications investigating the radiosensitising properties of gold nanoparticles, studying the impact of factors such a particle size, shape, and surface coating (2). These investigations made use of both mathematical modelling of their interactions with incident ionising radiation (3–7) as well as numerous *in vitro* and *in vivo* experimental studies (8–11).

Despite this, it remains an open question as to whether gold is the optimum material for this purpose. Only a handful of other elements have been investigated for use as radiosensitisers (including platinum, hafnium, gadolinium, and iron (12–16) ), and there have been no systematic experimental or theoretical comparisons between different materials.

The focus on gold largely stems from the original rationale for the use of high-Z contrast agents. If sensitisation derives from increased absorption, then it is reasonable to seek to maximize the nanoparticles' atomic number, as X-ray mass energy absorption coefficients increase strongly with increasing atomic number (with the photoelectric effect scaling as $Z^3$). Thus gold, being biocompatible and one of the heaviest stable elements, was a natural choice.

While these assumptions are known to be valid for imaging applications, which are driven primarily by the attenuation and absorption coefficients of the contrast agent, experimental studies of gold nanoparticle radiosensitisers have challenged this view for therapy. In this context, it is important to distinguish between *dose enhancement* – that is, the increase in energy deposited in the target volume due to the presence of the nanoparticles – and *radiosensitisation*, the increase in the biological effects of radiation observed when used in combination with nanoparticles. Although it was originally expected that these effects should be closely related for high-Z contrast agents, experimental investigations have shown that the radiosensitising effects of gold nanoparticles are very poorly correlated with macroscopic dosimetric calculations (2), with little or no relationship apparent between calculated dose enhancement and observed biological effects. In particular, radiosensitisation is often seen to be significantly greater than the increase in physical dose, and effects are seen using clinical megavoltage X-ray sources where the addition of nanoparticles leads to only negligible increases in the total dose. These results indicate that macroscopic dose enhancement alone is not a useful predictor of radiosensitisation across different cell lines and nanoparticle preparations.

As a result, several new hypotheses have been advanced to attempt to understand and predict these biological effects. One key observation from modelling of nanoparticle-radiation interactions is that, on the micro- and nano-scale, the dose distribution around gold nanoparticles is highly heterogeneous. Extremely high doses are deposited in the immediate vicinity of the nanoparticle, driven by the large number of low-energy secondary Auger electrons produced following ionisation in high-Z elements (17–19). Similar heterogeneous dose distributions are seen in ion-based radiotherapy, where techniques such as the Local Effect Model (LEM) have been developed to explain their superior biological effectiveness compared to relatively uniform X-ray exposures (20, 21). Analysis of nanoparticle-enhanced therapy based on these techniques have shown a similar increase in biological impact, potentially explaining some of the observed sensitisation in gold nanoparticle enhanced radiotherapy (22–24)

If nanoscale dose deposition is an important factor in the radiosensitising impact of nanoparticle contrast agents, then it is no longer clear that the heaviest elements are the best choice, as these nanoscale effects are poorly characterised by macroscopic dose, and care must be taken to incorporate the particles' potentially different impacts when exposed to X-ray imaging and therapeutic energies.

This work presents the first systematic computational study of the impact of elemental composition on nanoparticle-radiation interactions for both kilovoltage and megavoltage X-ray exposures, spanning elements from silicon (Z=14) to mercury (Z=80). While some of these elements may not be suitable for use as nanoparticle contrast agents, the full range of elements in this range was investigated to fully explore the underlying mechanisms of dose deposition.

For all nanoparticles, these simulations calculated total dose deposition, nanoscale dose distribution, and biological effects assessed through an approach based on the Local Effect Model. Complete reference data sets for these particles are also made available in the Supporting Information.

## Results

### Macroscopic analysis

Figure 1 illustrates the mass energy absorption coefficient for soft tissue and a series of high-Z materials (Gold, Hafnium, Gadolinium and Iodine) which are of interest as contrast agents, along with their ratio which is a guide to macroscopic dose enhancement. Although there is a general trend for higher Z atoms to see greater absorption at keV energies, this is not universally true as the positioning of the K and L absorption edges lead to sharp discontinuities which can often lead to lighter elements seeing stronger absorption. Thus even in the relatively simple macroscopic case, optimum contrast is not necessarily delivered by the material with the greatest atomic number.

This is further illustrated in Figure 2, which maps the relative potential dose enhancement per unit mass of contrast agent for monochromatic X-ray exposures. At low energies, these effects are dominated by bands representing elements which are strongly absorbing due to M, L, or K edge effects (respectively, from low to high energy). However, it can be seen that these bands are quite broad, with large numbers of elements within 25% of the maximum dose enhancement at a wide range of energies.

Higher energies see significantly less material dependence, as these effects are dominated by Compton interactions which are largely independent of atomic number, with all elements offering within 20% of the maximum contrast in the Compton dominated region from 1 to 5 MeV.

Notably, it can also be seen that all element with Z>=60 are within approximately a factor of 2 of the maximum achievable macroscopic contrast at all energies >= 10 keV. This suggests that even in imaging applications there is value in further investigating materials in this range, as the small increase in concentration needed to drive macroscopic effects may be offset by the greater flexibility offered by a wider range of candidate materials for nanoparticle design.

*Contributing processes on the macro- and nano-scale.*

A breakdown of the processes contributing to both total and local (within 1 μm) dose deposition is

shown in Figure 3, for 20 nm diameter nanoparticles exposed to either tuned kilovoltage irradiation

(left) or a 6 MV Linac spectrum at the 80% PDD (right). For kilovoltage interactions, the total energy deposition is dominated by photoelectrons and fluorescence photons at low and high energies respectively, with a small contribution of Auger electrons and Compton scatter. However, considering only dose deposited within 1 μm of the nanoparticle, Auger electrons are the primary source of energy deposition for all elements, as photoelectrons and fluorescence photons have too long a range to deposit significant dose in the vicinity of the nanoparticle.  Notably, unlike other processes Auger energy distribution has a multi-peaked behaviour on this scale, driven by the variation in Auger electron yield, energy and range as a function of atomic number.

In megavoltage exposures, total energy deposits are dominated by Compton scatter and there is now a significant contribution from secondary electrons generated by the beam interacting with the surrounding water volume, although the contributions of photoelectrons and fluorescence does increase for the heaviest elements. The distribution is again noticeably different on the local scale, where electron impact remains a dominant contribution but the effects of Compton electrons are mitigated due to their long range and a larger contribution of Auger electrons is seen, although still on a smaller scale than in kilovoltage exposures. Significantly, the total short-range energy deposit per interaction is roughly constant for MV exposures, as a result of electron scattering's weak dependence on atomic number.

*Nanoscale energy and dose distributions*

Figure 4 presents the nanoscale radial energy and dose distributions for an average ionising event

(that is, a radiation-nanoparticle interaction which produces at least one secondary particle) in nanoparticles of a selection of elements, under the same conditions as in Figure 3.

As expected from Figure 3, there is considerable variability among the different elements, driven by variations in Auger spectra. Differing Auger electron distributions can drive very high depositions in the immediate vicinity of the nanoparticle, broad peaks at moderate ranges (several hundred nm) or mixtures of these effects. These variations mean that it is challenging to predict which material delivers the highest dose enhancement at different distances from the particle, or which may offer the greatest radiosensitisation in general.

By contrast, megavoltage irradiations see only small variation with material. Again, this is in line with Figure 3, as the primary mechanisms of interaction (Compton and electron scattering) have very little dependence on atomic number for either cross-section or secondary electron spectra, giving broadly similar responses for all elements.

Because of the localised nature of interactions with these nanoparticles, for both cases these energy distributions correspond to very high local doses in the immediate vicinity of the nanoparticle, with kV exposures again showing significantly more variation than MV exposures.

*Biological Impacts*

An approach based on the Local Effect Model (LEM) was used to assess the potential biological impact of these dose distributions, as presented in Figure 5.  A complex material dependency is once again seen for kilovoltage exposures, with two distinct local maxima seen in the rates of damage predicted by the LEM, centred approximately around Z=34 and Z=68. Elements in these energy ranges have their primary Auger electrons (from K and L shell, respectively) with energies around 9 to 10 keV, with numerous additional lower-energy electrons with energies around 2 keV and below, corresponding to ranges in water on the order of 1 µm and 100 nm, respectively.

Material has a limited impact on MV Linac exposures – while there is a slight variation due to Auger electron contributions around Z=34 and 68, these are small, compared to the largely constant background independent of atomic number, driven by similar absorption of secondary electrons from the MV spectrum.

A small number of elements, including Europium, Gadolinium and Ytterbium, seem to lie significantly above the overall trend. A comparison of their physical properties suggests the common feature driving this effect is a relatively low density. This reduces self-absorption of secondary electrons, which can substantially increase the dose deposited in the vicinity of the nanoparticles.

Relative Biological Effectiveness (RBE) for nanoparticle-enhanced treatments in these conditions is presented in Figure 6, calculated for cells exposed to 500 µg/mL of uniformly dispersed 20 nm diameter nanoparticles exposed to a dose of 2 Gy. The trends with atomic number largely mirror those in Figure 5, with significant variation seen in keV energies but relatively limited variation for MV Linac exposures. RBE calculations also introduce a dependency on the relative ionisation cross-section, however, which acts to significantly reduce the impact of low-Z agents at keV energies, and all materials at MeV energies.

## *Discussion*

Studies of nanoparticle contrast agents and sensitisers have focused on very high-Z agents. While this follows naturally from the macroscopic dose calculations, an increasing body of evidence suggests this is a poor guide for biological radiosensitisation. Expanding this field to encompass other nanoparticle compositions may not only enable better tuning of nanoparticle dose distributions, but also the development of novel nanoparticle designs exploiting other element's physical, chemical or biological properties.

In this study we report on significant variations in predicted radiosensitisation between different elements which are not well described by macroscopic dose enhancement. These effects are driven by differences in Auger electron spectra, which depend primarily on the irradiated element. The contribution of Auger electrons at short range initially rises due to increasing energy deposited by K-shell Auger electrons before falling as their energies become too great to deposit significant local energy and their yield falls due to competition from fluorescence. However, at higher atomic numbers L- and M-shell electrons become sufficiently energetic to contribute to radiosensitisation, leading to a peak around Z=68. Finally, at the upper limit of this study L-shell Auger electron energies also become too great and their local contribution begins to reduce. Alongside these trends is a variation in the range over which energy is primarily deposited, as can be seen in Figure 4. This may be an important factor in determining sensitising properties if nanoparticles are not uniformly distributed, as has been assumed in this analysis.

Interestingly, for a clinical megavoltage source, Auger electrons are significantly less important across all elements, as interactions are dominated by Compton and electron scattering events, which are primarily outer-shell interactions. This leads to a relatively material-independent prediction for sensitisation at megavoltage energies, and significantly lower overall effect.

This observation may prove very important for the usage of particles in a theranostic context, where imaging and therapeutic functions are combined. The very different interactions at high and low energies may present challenges in these applications, such as greater than expected sensitisation in normal tissues to pre-treatment imaging delivered by CT in a contrast-enhanced setting. As a result, care must be taken to evaluate the interactions of these particles with all aspects of the treatment pathway.

These results also present an avenue for validating models of the biological impacts of nanoparticle radiosensitisers. As the nanoscale dose model suggests a specific, complex dependence of radiosensitisation on nanoparticle material, comparing the radiosensitising properties of nanoparticles composed of different materials in biological systems offers a sensitive probe of the validity of these assumptions, in comparison to other possible radiosensitising mechanisms such as biological or chemical stresses induced by the nanoparticles which would be expected to have a different material dependence. Such approaches of course depend on the development of nanoparticles with different compositions but similar biological uptake and localisation.

If validated, this wider range of material choices suggested by this analysis may open novel options for new nanoparticle designs which may make use of a range of elements in their design. This has the potential to offer more affordable therapeutic options, as well as new methods for optimisation of sensitisation, taking advantage of the chemical or biological properties of materials which may initially have been rejected as poor candidates for radiosensitisation due to their low atomic number. As noted above, this is also potentially significant for the development of theranostic nanoparticles, whether directly through improved X-ray absorption within the target, or by selecting elements which are useful for alternative imaging techniques, such as MRI.

One other observation in these results is the contribution of particle density to sensitisation. The least dense elements produce dose distributions which are predicted to lead to significantly greater sensitisation than than would be expected based on atomic number alone, which is believed to be driven by low-energy secondary electrons having an increased probability of escape. This suggests that less dense particle preparations (e.g. combining some atoms of a high-Z material in a crystal or organic molecule with lighter elements) may drive superior radiosensitisation than a similar mass of material contained in denser pure nanoparticles (subject to the ability to deliver a sufficient total concentration of contrast agent). Such an approach has been taken in development of hafnium-oxide and gadolinium-based nanoparticles, which have reported significant radiosensitising properties (13, 14) and are moving towards clinical trials.

It is important to note that this is a preliminary exploration of the impact of material choice in nanoparticle radiosensitisation, only considering pure elemental nanoparticles of one size and limited energy selections to illustrate underlying mechanisms. Specific real exposure conditions are likely to differ, with different irradiation energies leading to changes in interaction cross-sections, and nanoparticle size having a significant impact on the low-energy portion of the secondary electron spectra.

Significant material-specific tuning is likely to be possible, taking into greater account the particular characteristics of its secondary electron energy deposition and particle design. Additionally, a detailed optimisation of nanoparticle design would need to investigate not only the physical properties of these materials but also their biological and chemical comparability and any associated coatings, including how this impacts on both the quantity and uniformity of nanoparticle delivery. However, the current analysis is still sufficient to demonstrate that there is a viable physical rationale not to focus exclusively on the heaviest elements when developing radiosensitising nanoparticles.

## *Conclusion*

In conclusion, these results present a much more complex picture of the radiosenisiting properties of heavy atom nanoparticles than would be expected from their mass energy absorption coefficients alone. This suggests that there is considerable merit in investigating nanoparticle preparations which make use of other light elements, to potentially optimize the radiation-related sensitising effect and to make best use of the wide range of chemical and biological properties which would be accessible through novel nanoparticle designs.

## *Experimental Section*

### *Analytic macroscopic dose enhancement calculations*

To provide a reference against which to compare the nanoscale calculations, macroscopic dose enhancement factors were calculated for all elements. Based on the assumption that nanoparticle contrast agents can be considered as homogeneously distributed throughout a target volume, potential dose enhancement per unit mass of contrast agent can be approximated simply as the ratio of the mass energy absorption coefficients, $DER(Z) = \frac{\mu_Z}{\mu_{tissue}}$ . This can then be scaled by the particle concentration to give an actual dose enhancement value. These values were calculated for all materials as a function of radiation energy based on values from NIST (25).

For ease of comparison between different materials, a normalised ratio has been calculated at each energy, normalising each material's dose enhancement ratio to the maximum contrast possible at that energy. This offers an easier comparison between different materials, as the maximum achievable dose enhancement varies greatly as a function of energy.

*Monte Carlo nanoparticle dose deposition calculations*

Radiation-nanoparticle interactions and resulting radial dose distributions were modelled using Geant4.9.6 (patch 3) (26), simulating individual 20 nm diameter nanoparticles placed within the centre of a 10 µm a side cube of water. Livermore low-energy physics models were used for radiation transport within the nanoparticle volume, with Geant4-DNA models used in the surrounding water volume (27). The use of water as a detector volume for the dose distribution is necessarily an approximation to biological systems which contain a wider range of different chemical species, but is a necessary simplification due to the current lack of appropriately detailed models of low-energy radiation interactions with organic systems.

Nanoparticles were modelled as pure spheres of individual elements ranging from Z=14 to Z=80, with material properties (isotope distribution, density, etc) taken as Geant4 defaults for STP, based on the NIST reference values. Elements which were liquid or gaseous under these conditions were not considered for further analysis due to poor statistics.

Initially, interactions were modelled using monochromatic keV X-rays. As ionisation cross-sections and Auger spectrum depend strongly on photon energy for keV X-rays, individual X-ray energies were used for each material. Energies were set to 20 keV above the K-edge of the material being exposed (ranging from 22 to 102 keV). This enables comparisons to be made between all elements following similar ionising events, which primarily occur in inner shells to allow for the full impact of Auger electrons to be investigated. Total primary particles simulated ranged from 8×108 to 1.6×109 photons, depending on nanoparticle and beam energy, delivered as a 20 nm diameter beam exposing the whole particle.

Clinically-relevant Megavoltage exposures were modelled using both the photon and electron components of a 6 MV Linac spectrum at the 80% PDD, obtained through Monte Carlo simulation as described previously (23, 28). In these simulations, 3×108 primary photons sampled from a published 6 MV Linac spectrum (29) were directed in a 10 cm diameter beam along the axis of a cylindrical block of water of 20 cm length and diameter. These simulations scored dose along the beam direction as well as the spectrum of both photons and electrons at the 80% dose-depth position, 8.7 cm below the surface. To model nanoparticle irradiations, these spectra were scaled down to nanometer scales and used to expose individual nanoparticles, as was the case for kilovoltage photons. Once again, radial doses and secondary particle distributions were calculated for nanoparticles with atomic numbers ranging from 14 to 80. These simulations modelled 1.6×109 photons, and approximately 2.5×107 electrons, based on the input phase space, scaled to a 20 nm diameter beam.

For both types of exposure, all secondary particles emitted from the nanoparticles were scored, identifying the process which led to their emission, as well as the dose deposited in concentric 2 nm shells around the nanoparticle, out to a range of 1 μm from the nanoparticle.

For each nanoparticle/radiation combination studied in this work, the full secondary particle distributions as well as the radial dose distributions (broken down by contributing process) are provided in the Supplementary Information to support further investigation.

Nanoscale Radiosensitisation Calculations

An approach based on the Local Effect Model (LEM-1) was used to evaluate the potential biological impact of these nanoscale dose depositions. The LEM was developed to describe the biological effectiveness of highly charged ions (20, 21), which have a significantly higher biological effectiveness than a similar dose of X-rays. This technique, as well as its application to nanoparticle-enhanced therapy, has been described in detail elsewhere (20, 23), and is reviewed below for completeness.

The LEM suggests that the biological effectiveness of heterogeneous exposures can be understood in terms of the dose at each point within the cell, rather than the average dose to the cell. Specifically, it postulates that cells die to the formation of 'lethal lesions', with survival given by $S(N) = e^{-N}$, where N is the number of lesions formed in the cell. N can be expressed as $N(D) = \ln(S\{D\})$, where S(D) is the survival of cells following exposure to a uniform dose D of X-rays.

For heterogeneous exposures, the LEM assumes that a) the microscopic lesion density at given dose is the same as that across the whole cell; and b) the total number of lesions within the cell is dependent on the integral of the probability of a lesion forming at each point. Thus, the total number of lesions is given by $N_{tot} = \int N(D_r) \frac{dV}{V}$, where $D_r$ is the dose delivered to the point r, exposing a fraction of the total cell volume given by $\frac{dV}{V}$. As a result of non-linear terms in the dose response function, the localised doses of highly charged ion therapies drive significantly more cell killing than a uniform exposure to the same average dose of X-rays.

These concepts can also be applied to nanoparticle-enhanced therapies, and have successfully demonstrated more accurate prediction of radiation sensitization in nanoparticle enhanced therapies, suggesting these effects may play an important role in nanoparticle sensitization (22, 23).

In this work, the number of lesions induced per nanoparticle-radiation interaction, $N_{NP}$, was calculated by summing the damage as a function of distance from the nanoparticle, assuming the linear-quadratic model of response to uniform exposures, $S(N) = e^{-\alpha D - \beta D^2}$. This gives a total number of lesions as $N_{NP} = \int (\alpha D_{NP}(r) + \beta D_{NP}(r)^2) dV / V$, where Dnp(r) is the radial dose distribution around a nanoparticle following an ionising event (as illustrated in Figure 4). For ease of comparison, the potential cell killing impact of a single nanoparticle-radiation interaction was equated to a uniform X-ray dose $D_{Eff}$, defined as $N_{NP} = \alpha D_{Eff} + \beta D_{Eff}^2$.

For the purposes of this analysis, the calculated radial dose distribution is smoothed using a spherically symmetric Gaussian kernel, as in ion therapy approaches (20), with σ=10 nm. This represents the diffusion of potentially damaging biological and chemical species following initial radiation interactions while preserving the total energy deposited in the system. This has several effects, including a reduction in sensitivity to statistical uncertainties and the removal of non-physical dose peaks at extremely small radial positions.

Finally, these effects are converted into a Relative Biological Effectiveness (RBE). The RBE is defined as $RBE = \dfrac{D_X}{D_{NP}}$ where Dx is the reference X-ray dose, chosen to be 2 Gy in this work, and DNP is the dose which yields equal survival in the presence of nanoparticles. The number of lethal lesions (and thus survival) was calculated for a given condition according to

$N_{tot} = N_X + \eta N_{np}$, where $N_X$ is the number of lesions induced by the uniform X-ray dose, $N_{NP}$ is the number of lesions induced per nanoparticle-radiation interaction, and $\eta$ is the number of nanoparticle-radiation interactions in a given exposure. $\eta$ is calculated as the product of the delivered dose, the number of interactions per nanoparticle per Gray for the particle/radiation type under consideration (taken from ionisation cross-sections for monoenergetic keV exposures, and the phase space calculations for linac exposures) and the number of particles present in the volume, taken to be 500 µg/mL in this work.

It is important to note that analysis assumes that all of the volume around the nanoparticle are uniformly sensitive to ionising radiation. While this may not always apply for nanoparticles which may, for example, be sequestered far from DNA and other sensitive targets by active cellular processes, it is a useful initial guide to sensitising impacts.

## Supporting Information

The supporting information contains simulation results for both keV and Linac exposures used to generate data presented in this article. Full sets of data are available containing radial energy deposit distributions per ionising event, secondary particles per ionising event, and total event counts for simulations.

## Acknowledgements

The authors are grateful to the European Commission Framework 7 Programme (grant numbers EC FP7-MC IOF-623630 – RadResPRO and EC FP7 MC-ITN-608163 – Argent) for funding their work.

## *Figures*

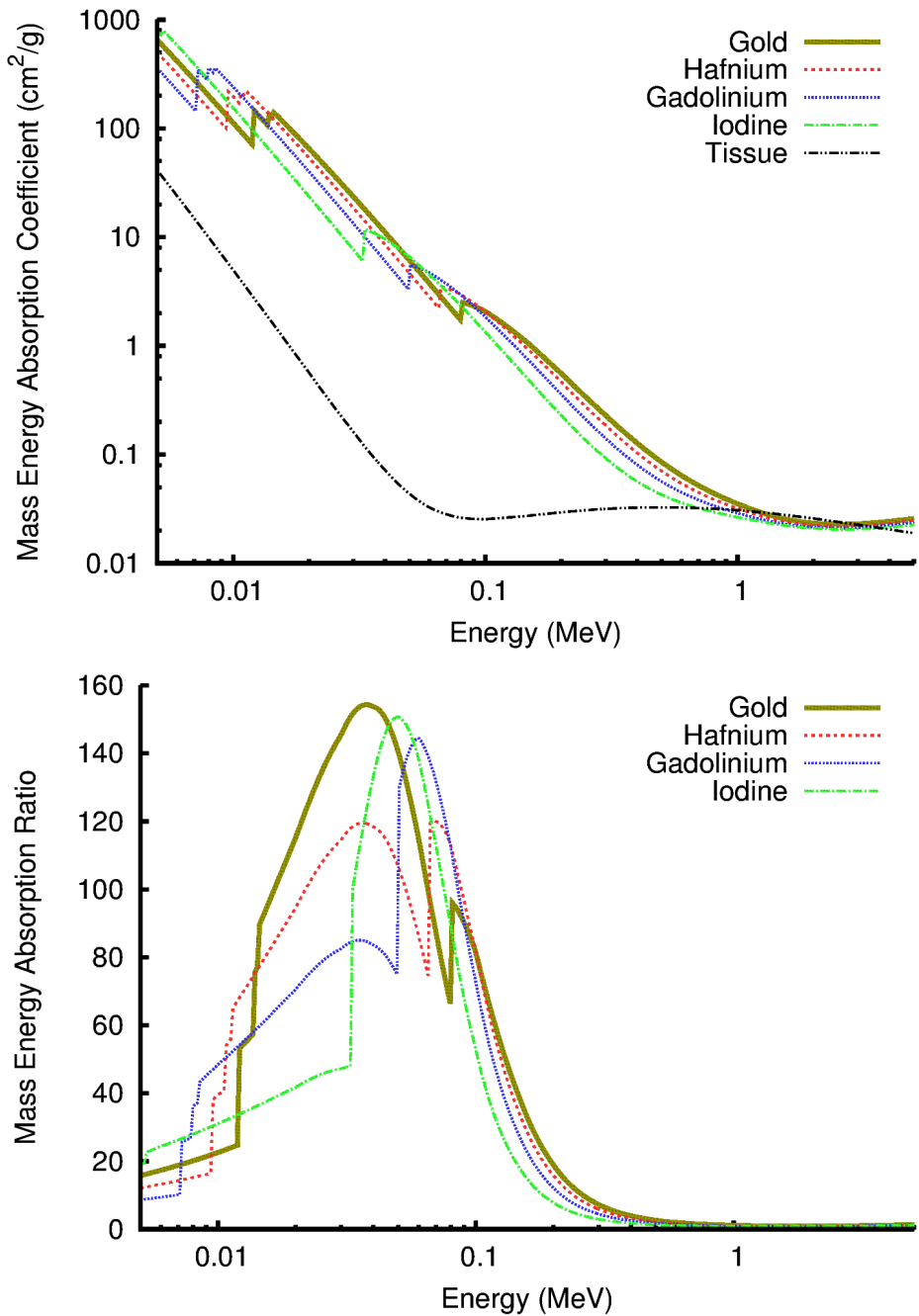

**Figure 1:** Comparison of Mass Energy Absorption coefficients for soft tissue and a range of heavy elements (top). Although higher-Z metals generally have the highest absorption coefficient, this is not always the case, with edge structure introducing significant variation. This is similarly apparent in the ratio of metal absorption to that of soft tissue (bottom), which shows gold's absorption is surpassed by other metals over a wide range of the kilovoltage region.

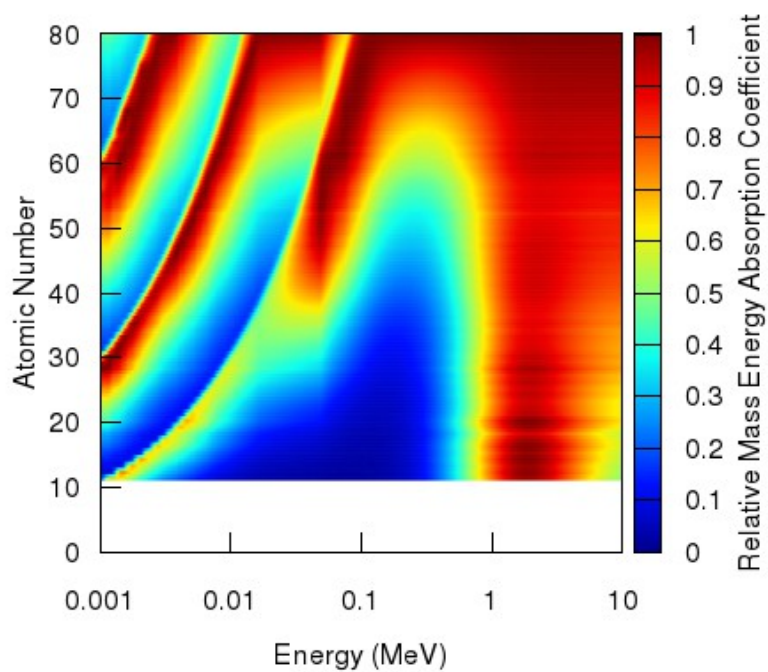

**Figure 2:** Normalised enhancement per unit mass for a range of materials and energies. At each energy, the enhancement ratio has been normalised to that of the maximum at that energy. Clear structure can be seen at low energies, with bands corresponding to K-, L- and M-shell absorption. At higher energies, little variation in potential enhancement is seen as absorption is dominated by Compton interactions.

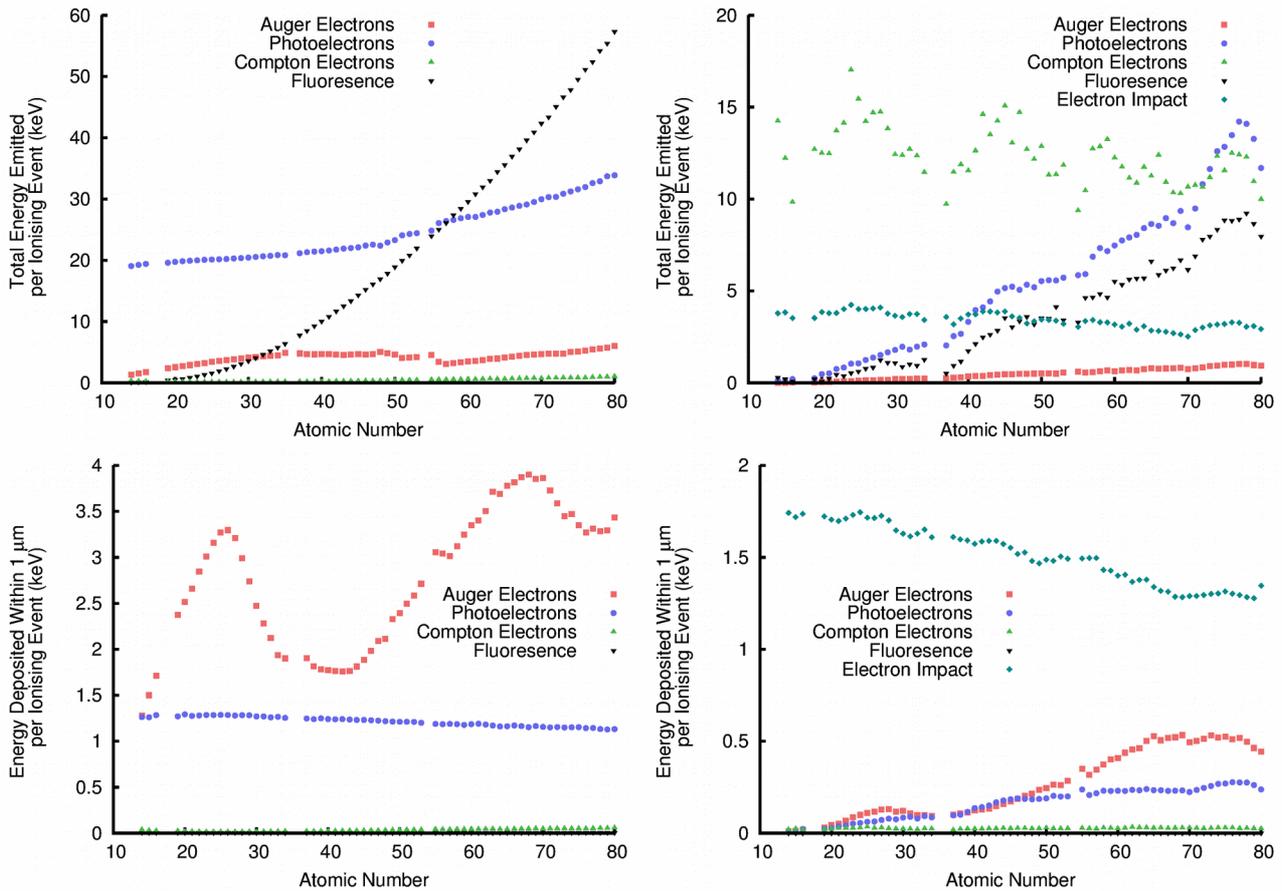

**Figure 3:** Top: Total energy emitted from a nanoparticle by various processes following an ionising event as a function of atomic number, for tuned keV irradiation (left) or 6 MV Linac exposure (right). At keV energies, photoelectrons and fluorescence dominate these effects, while for MV energies ionisations by secondary electrons in the beam spectrum dominate. Bottom: Distribution of energy deposited within 1 micron of a nanoparticle centre per ionising event, broken down according to contributions of various processes. At keV energies, the majority of energy is deposited by Auger electrons, which have a complex energy dependence, while for MV interactions electron impact remains the dominant contribution.

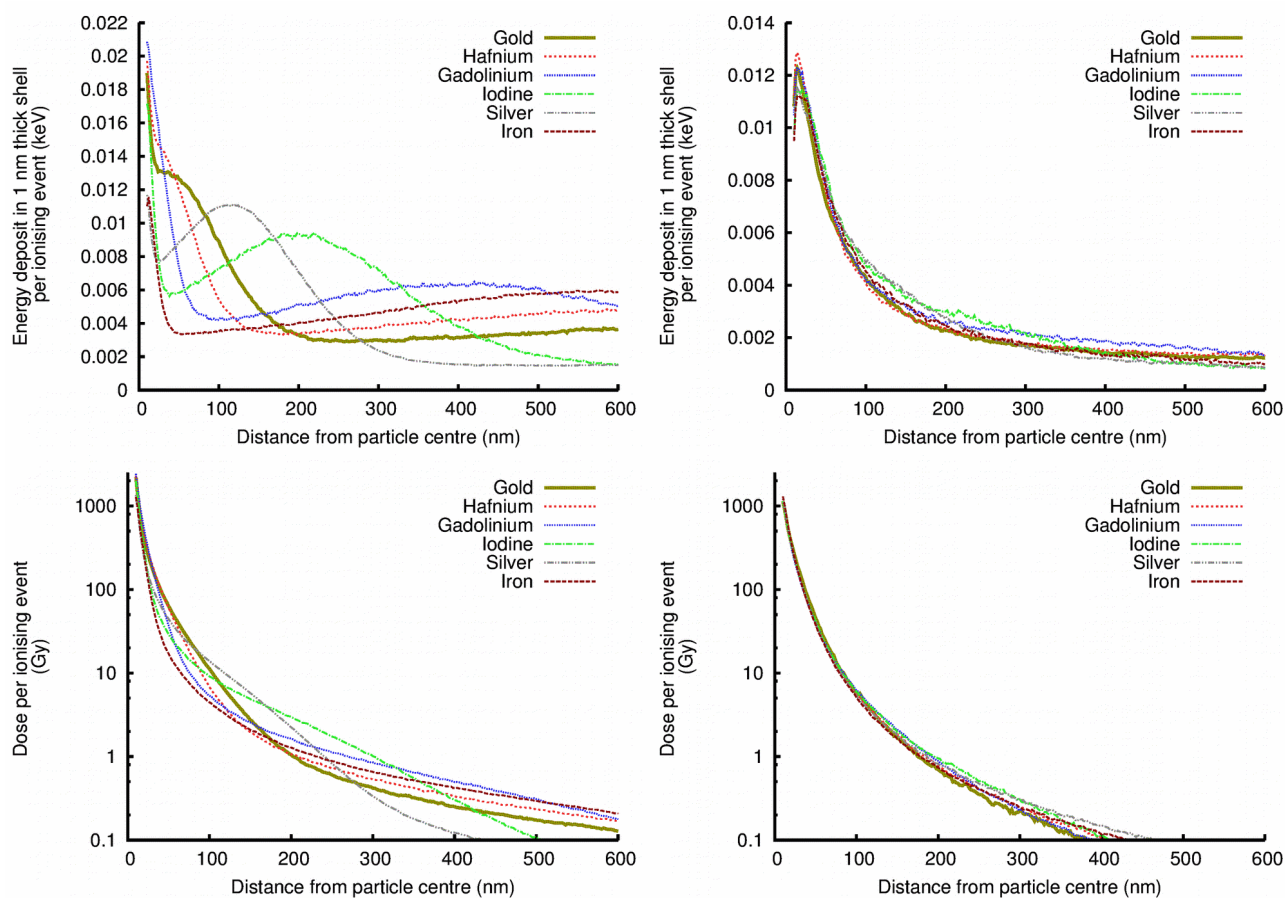

**Figure 4:** Nanoscale radial energy distributions (top) and dose distributions (bottom) around 20 nm nanoparticles of various metals following a single ionising event caused by either tuned keV X-rays (left) or a 6 MV Linac spectrum (right). Significant complexity is seen in keV irradiations, with differing energy distributions depending on the characteristic Auger cascade produced by the material. By contrast, MV irradiation produces similar energy distributions in all cases, as Auger electrons play a reduced role. Bottom: For both spectra, there remain high localised doses in the vicinity of high-Z nanoparticles.

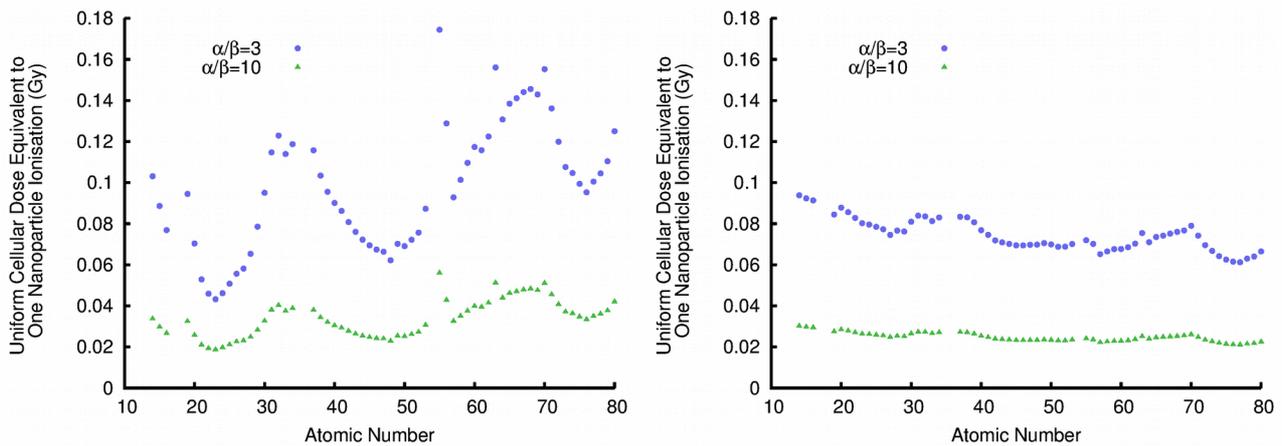

**Figure 5:** Biological effects of nanoparticle-radiation interactions, in terms of effective dose deposited by a single ionising event, for cells with α/β ratios of either 3 (blue) or 10 (green) for keV (left) and MV (right) exposures. For keV irradiation, significant variation is seen, with several distinct peaks of effect, reflecting nanoscale Auger dose distributions. By contrast, MV interactions are relatively slowly-varying over the entire range of atomic number.

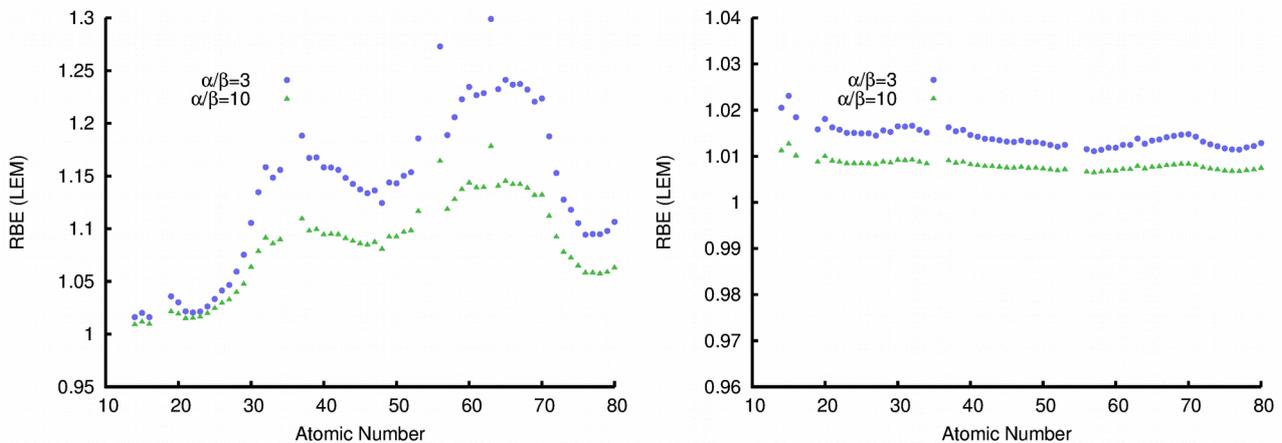

**Figure 6:** Relative Biological Effectiveness (RBE) predicted by the LEM for addition of nanoparticles to cells at concentrations of 500 μg/mL for 2 Gy exposures using kilovoltage (left) and megavoltage (right) exposures. These trends largely follow those seen in in the per-interaction rates in Figure 5, but there is also a significant contribution from the interaction cross-section of the radiation, reducing the impact of the lightest elements at keV energies, and all elements at MeV energies.